\documentclass[prc,showpacs,aps,twocolumn]{revtex4}

\usepackage{amsmath}

\usepackage{epsfig}

\begin{document}

\bigskip

\title{Threshold  Effects in Multi-channel Coupling and Spectroscopic
Factors in Exotic Nuclei}

\author
{N. Michel$^{\dag \ddag \P}$, W. Nazarewicz$^{\dag \ddag \$}$, 
M. P{\l}oszajczak$^{\S}$}

\affiliation{$^{\dag}$ Department of Physics and Astronomy, University of Tennessee, Knoxville, TN 37996, USA\\
$^{\ddag}$ Physics Division, Oak Ridge National Laboratory, P.O.~Box 2008, Oak Ridge, TN 37831, USA\\
$^{\P}$ Joint Institute for Heavy Ion Research, Oak Ridge, TN 37831, USA\\
$^{\$}$ Institute of Theoretical Physics, Warsaw University, ul.~Ho\.{z}a 69, 00-681 Warsaw, Poland\\
$^{\S}$ Grand Acc\'el\'erateur National d'Ions Lourds (GANIL), CEA/DSM - CNRS/IN2P3,
BP 55027, F-14076 Caen Cedex, France}

\date{\today}

\begin{abstract}

In the threshold region,  the cross section and the associated overlap
integral obey the Wigner threshold law  that results in the Wigner-cusp
phenomenon. Due to  flux conservation, a cusp anomaly in one  channel
manifests itself  in other  open channels, even if their respective
thresholds appear at a different energy. The shape of a threshold cusp
depends on the  orbital angular momentum of a scattered particle; hence,
studies of Wigner anomalies in weakly bound nuclei with several
low-lying thresholds can provide valuable spectroscopic information. In
this work,  we investigate  the threshold behavior of  spectroscopic
factors in neutron-rich drip-line nuclei using the Gamow Shell Model,
which takes into account many-body correlations and  the continuum
effects. The presence of threshold anomalies is demonstrated and the
implications for spectroscopic factors are discussed.

\end{abstract}


\bigskip 

\maketitle

In 1948, based on general principles, Wigner predicted \cite{Wigner} a
characteristic behavior (a cusp) of  scattering and reaction cross
sections in the vicinity of  a reaction threshold. This particular
behavior (often referred to as the Wigner's threshold law or the
Wigner-cusp phenomenon) was given a quantitative explanation a decade
later \cite{Breit,Baz,Newton,Meyerhof,Baz1,Lane}. In particular, it has
been noted \cite{Baz,Newton,Meyerhof,Baz1,Hategan1} that, due to the
unitarity of the scattering matrix and  the resulting flux conservation,
the presence of  a threshold anomaly in an opening reaction channel can
trigger an appearance of Wigner cusps in other open  channels with lower
thresholds. As the shape of the cusp strongly depends on orbital angular
momentum (is strongest in $s$ and $p$ waves), it was early realized that
the presence of cusp anomaly could provide structural information about
reaction products  \cite{Baz,Newton,Adair}.

The Wigner-cusp phenomenon has been studied experimentally and
theoretically in various areas of physics: pion-nucleus scattering
\cite{Adair,Starostin}, electron-molecule scattering \cite{Domcke},
electron-atom scattering  \cite{Scheibner}, and ultracold atom-diatom
scattering \cite{Forrey}. In low-energy nuclear physics, threshold
effects have been investigated in, e.g.,  charge-exchange reactions
\cite{Malmberg}, neutron elastic scattering \cite{Wells}, and deuteron
stripping \cite{Moore}. Abramovich {\it et al.} \cite{Abramovich}
reviewed threshold phenomena in nuclear reactions, including those with
the light neutron-rich  systems such as $^6$He, $^{10}$Be,  and
$^{10}$Li (see also Ref.~\cite{Hategan}) that are of particular interest
in the context of this work. The influence of threshold effects on 
the cross section and the strength function, hence
the spectroscopic factor (SF)  of a threshold state was 
pointed out in Refs.~\cite{Hategan1,Graw}.

The purpose of this study is many-fold. Firstly, we investigate whether
the Wigner-cusp phenomenon appears naturally in a microscopic many-body
approach rooted in an effective inter-nucleon interaction. The second
goal is to investigate the influence of threshold effects and the
coupling to the non-resonant continuum on overlap integrals, or
spectroscopic factors. We demonstrate  that the energy dependence of SFs
caused by an opening of a reaction channel can be described at the
shell-model level only if nucleon-nucleon correlations involving
scattering states are treated properly. Finally, we emphasize the
importance of experimental studies of cusp phenomena in weakly bound,
neutron-rich nuclei in which  low-lying one-neutron and two-neutron
thresholds appear.

The traditional shell model (SM) of nuclear structure views  the nucleus
as a  closed quantum system (CQS) in which nucleons occupy bound orbits.
While such an  assumption may be somehow  justified for well-bound
nuclei  having  high  particle-emission thresholds, it can no longer
be applied to weakly bound or unbound systems in the vicinity of
drip lines, where the coupling to the particle continuum (both
resonances and the non-resonant scattering states) becomes important.
This  coupling  can be considered in the open quantum system (OQS)
extension of the SM, the so-called continuum SM (see Ref.
\cite{opr} for a recent review). In this work,   we apply the
complex-energy implementation of the continuum SM, the so-called Gamow Shell Model
(GSM) \cite{Mic02,Bet02} in the version of Refs.~\cite{Mic02,Mic04}. GSM
is a multi-configurational SM with a single-particle (s.p.) basis given
by the Berggren ensemble \cite{Berggren1} which consists of Gamow (bound
and  resonance) states and the  non-resonant continuum. For a given
Hamiltonian, the number of particles occupying states of the
non-resonant continuum is a result of GSM variational calculations. The
resonant states of the GSM are the generalized eigenstates of the
time-independent Schr\"{o}dinger equation which are regular at the
origin and satisfy purely outgoing boundary conditions. The GSM can thus
be viewed as  a quasi-stationary many-body OQS formalism.
(An alternative microscopic approach, successfully applied to the structure of
weakly bound or unbound nuclei, is the Complex Scaling Method 
that  is capable of treating different kinds of reaction channels and continuum
states starting from different thresholds \cite{Myo}.)
Since the Wigner-cusp phenomenon is most pronounced for  low-$\ell$
waves and for neutrons (no Coulomb barrier), as an illustrative example
we choose the case of the one-neutron (1n) channel in  the model two-
and three-neutron systems outside the inert core:  $^{6}$He and $^7$He.
Our aim is not to fit the actual experimental data, but rather to
accomplish the physics goals as  stated above. 

\paragraph*{Gamow Shell Model Framework--}
The starting point of GSM is the Berggren one-body completeness relation
\cite{Berggren1} allowing the expansion of both bound and unbound states.
The Berggren ensemble consists of resonant and scattering states generated
by a finite-depth potential $V(r)$. 
Resonant states are solutions of the Schr\"{o}dinger equation with
purely outgoing asymptotics. Their energies and wave functions are in
general complex. The resonant states of the GSM have either bound or
decaying character; they  form the so-called  {\it pole subspace}.
Scattering states entering the Berggren ensemble  are also defined in the complex
energy/momentum plane. For a given partial wave $(\ell,j)$, the
scattering states are distributed along a contour $L_+^{\ell_j}$ in the
complex momentum plane. The set of all bound and decaying states $|u_n
\rangle$ enclosed between $L_+^{\ell_j}$ and the real $k$-axis, and
scattering states $|u_k \rangle$ in $L_+^{\ell_j}$ is complete
\cite{Berggren1}:
\begin{equation}
\int\hspace{-1.4em}\sum_{\mathcal{B}} |u_\mathcal{B} \rangle 
\langle \widetilde{u_\mathcal{B}}| = 1,  \label{one_body_compl_rel}
\end{equation}
where $|\mathcal{B}\rangle$ is either a discrete resonant state
or a scattering continuum state ($L_+^{\ell_j}$ part).
The tilde symbol above bra vectors in Eq.~(\ref{one_body_compl_rel})
signifies that  the complex conjugation arising in the dual space
affects only the angular part and leaves the radial part unchanged
\cite{Berggren1}. The continuous part of the completeness relation 
(\ref{one_body_compl_rel})  has to be discretized in numerical
applications. For that purpose, scattering
states are discretized and renormalized in order to obtain a discrete 
completeness relation \cite{Berggren1}:
\begin{eqnarray}
\sum_{\mathcal{B}=1}^{N} |u_{\mathcal{B}} \rangle \langle \widetilde{u_{\mathcal{B}}}| \simeq 1 
~~;~~~~|u_{\mathcal{B}} \rangle = \sqrt{\omega_{\mathcal{B}}} \; 
|u_{k_{\mathcal{B}}} \rangle, 
\label{one_body_compl_rel_discr}
\end{eqnarray}
where $\{k_{\mathcal{B}},\omega_{\mathcal{B}}\}$ is the set of
discretized momenta and associated weights provided by a Gauss-Legendre
quadrature. The many-body GSM basis corresponds to Slater
determinants (SD) spanned 
by  one-body Berggren states:
$\displaystyle |SD_i \rangle = | u_{i_1} \cdots u_{i_A} \rangle$ where
$|SD_i \rangle$ is the $i$-th SD 
in the $A$-body basis and $u_{i_j}$ is the $j$-th one-body state
occupied in $|SD_i \rangle$. The many-body completeness 
relation is built from Eq.~(\ref{one_body_compl_rel_discr})
by forming all possible SDs generated by the Gamow one-body states:
\begin{eqnarray}
&&\sum_{i} |SD_i \rangle \langle \widetilde{SD_i}| \simeq 1. \label{Nbody_compl_rel_discr}
\end{eqnarray}
The completeness in (\ref{Nbody_compl_rel_discr}) is not exact as the
one-body completeness relation (\ref{one_body_compl_rel_discr})
is approximate due to the discretization. 
In the basis (\ref{Nbody_compl_rel_discr}), the GSM Hamiltonian $H$
becomes a complex symmetric matrix.  Moreover, many-body bound and
resonant states are embedded in the background
of non-resonant
scattering eigenstates, so that one
needs a criterion to isolate them. The
overlap method \cite{Mic02} has proven to be very  efficient to solve
this problem. For this, one diagonalizes first $H$ in the pole
subspace  to
generate a zeroth-order vector  $|\Psi_0\rangle$. In the second
step, $|\Psi_0\rangle$ is used as a pivot to generate a Lanczos subspace
of the full GSM space. Its diagonalization provides
eigenvectors of $H$ in the total GSM space, and the requested bound or
decaying eigenstate of $H$ is the one which maximizes the overlap $|\langle \Psi_0 |
\Psi\rangle|$.

The definition of observables in GSM follows directly from the
mathematical setting of quantum mechanics in the rigged Hilbert space
rather than  the usual Hilbert space \cite{Bohm,Madrid}. Modified
definition of the dual space, embodied by the tilde symbol above bra
states, implies that observables in many-body resonances become complex.
In this case, the real part of a matrix element corresponds to the
expectation value,  and the imaginary part can be interpreted as the
uncertainty in the determination of this expectation value due to the
possibility of decay of the state during the measuring process
\cite{Berggren1,Civ99}.

\paragraph*{GSM Hamiltonian--}
In this Letter, the s.p.~basis (\ref{one_body_compl_rel}) 
is generated by a Woods-Saxon (WS) potential with the radius $R_0$=2 fm,
the diffuseness $d$=0.65 fm,  the
spin-orbit strength $V_{\rm so}$=7.5 MeV, and  the depth of the central
potential $V_0$=47 MeV (the ``$^{5}$He" parameter set). This potential
reproduces experimental energies and widths of the s.p.~resonances
$3/2_1^-$ and $1/2_1^-$ in $^5$He. The GSM Hamiltonian is a sum of the
one-body WS potential, representing the effect of an inert $^4$He core,
and of the two-body interaction among valence particles, given by a
finite-range surface Gaussian interaction (SGI) \cite{Mic04} with the
range $\mu=1$\,fm and the coupling constants depending on the total
angular momentum $J$ of the neutron pair: $V_0^{(0)}=-403$ MeV fm$^{3}$
and $V_0^{(2)}=-392$ MeV fm$^{3}$. These constants are fitted to
reproduce the experimental ground state (g.s.) binding energies of
$^6$He and $^7$He with the ``$^{5}$He" WS potential. The valence space
for neutrons consists of the $p_{3/2}$ and $p_{1/2}$ partial waves. The
$p_{3/2}$ wave functions include a $0p_{3/2}$ resonant state and
$p_{3/2}$ non-resonant scattering states along a complex contour
enclosing the $0p_{3/2}$ resonance in the complex $k$-plane. For a
$p_{1/2}$ part, we take non-resonant scattering states along the
real-$k$ axis (the broad $0p_{1/2}$ resonant state  plays a negligible
role in the g.s.~wave function of $^{6}$He). For both contours, the
maximal momentum value is $k_{\rm{max}}=$3.27 fm$^{-1}$. The contours
have been discretized with up to 60 points and the attained precision on
energies and widths is better than 0.1 keV. In the subsequent analysis,
parameters of the GSM Hamiltonian are varied in order to change
positions of 1n thresholds in various isotopes.

In order to illuminate the continuum coupling effects in GSM, we have
introduced  a simplified harmonic oscillator SM  (HO-SM) scheme. Here, 
the radial wave functions are those of the spherical harmonic oscillator
  with the frequency $\hbar \omega=41A^{-1/3}$ MeV. The one-body part of
the HO-SM Hamiltonian is given by the real energies of the one-body part
of the GSM Hamiltonian.  The HO-SM scheme is supposed to illustrate the
``standard"  CQS SM calculations in which only bound valence  shells are
considered in the s.p. basis.

\paragraph*{Spectroscopic Factors in GSM--} SFs are useful indicators of
the configuration mixing in the many-body wave function. Extensive
attempts have been made to deduce SFs using direct reactions, such as
single-nucleon transfer, nucleon knockout,  and elastic break-up
reactions, using hadronic and leptonic probes. These analyses often
reveal model- and probe-dependence \cite{sf1,sf2,sf3} raising concerns
about the possibility of their precise experimental determination. (For
an extensive study of spectroscopic factors in exotic nuclei from
nucleon-knockout reactions, see Ref.~ \cite{gade}.) Studies of $(e,e'p)$
reactions in closed-shell nuclei \cite{eep} demonstrated that the SFs
are reduced by $\sim 35\%$ with respect to the standard SM predictions,
mainly due to the coupling to high-momentum states reached by the
short-range and tensor components of the nucleon-nucleon interaction
\cite{dickhoff}. In this work, we point out that additional difficulties
in extracting and interpreting SFs from the measured cross sections
using the standard SM lie in  the neglect of particle continuum, channel
coupling, and strength fragmentation. 
 
Single-nucleon overlap integrals and the associated spectroscopic
factors (SFs) are basic ingredients of the theory of direct reactions
(single-nucleon transfer, nucleon knockout, elastic break-up)
\cite{satch,sfs}.   Experimentally, SFs  can be  deduced
from measured cross sections; they are useful measures of  the configuration mixing
in the many-body wave function. The associated reaction-theoretical analysis 
often reveals
model- and probe-dependence \cite{sf1,sf2,sf3} raising concerns about
the accuracy of  experimental determination of SFs. In our study we
discuss the  uncertainty in determining SFs due to  the two assumptions commonly used in  the
standard SM studies, namely (i) that a nucleon is transferred
to/from a specific s.p.~orbit (corresponding to an observed
s.p.~state), and  (ii) that the transfer to/from the continuum 
of non-resonant scattering states
can be disregarded.
In this work, we define SFs in a usual way, through
the radial
overlap  functions $u_{\ell j}(r)$  
\cite{sfs,Glenden,Fro_Lipp}:
\begin{equation}
u_{\ell j}(r) = \langle \Psi^{J_{A}}_{A} | \left[ | \Psi^{J_{A-1}}_{A-1} 
\rangle \otimes |\ell,j \rangle \right]^{J_A} \rangle, \label{ovf_basic_def}
\end{equation}
where $|\Psi^{J_{A}}_{A}\rangle$ and
$|\Psi^{J_{A-1}}_{A-1}\rangle$ are wave functions of nuclei $A$ and $A-1$,
respectively, and 
$|\ell,j\rangle$ is the angular-spin part of the channel function.
The angular-spin degrees of freedom
are integrated out in Eq.~(\ref{ovf_basic_def})
so that  $u_{\ell j}$ depends only on 
the relative radial coordinate of the transferred particle
$r = |\vec{r}|$.
The spectroscopic factor, denoted by $S^2$, is defined as usual
through the norm of
the overlap function \cite{sfs,Glenden,Fro_Lipp} . 
Using a decomposition of the $(\ell,j)$ channel function in the complete
Berggren basis,  one obtains:
\begin{eqnarray}
&&u_{\ell j}(r) = \int\hspace{-1.4em}\sum_{\mathcal{B}} \langle \widetilde{\Psi^{J_{A}}_{A}} || a^+_{\ell j}(\mathcal{B}) || \Psi^{J_{A-1}}_{A-1} \rangle
\mbox{ } \langle r \ell j | \mathcal{B} \rangle, \label{ovf_GSM} \\
&&S^2 = \int\hspace{-1.4em}\sum_{\mathcal{B}} \langle
 \widetilde{\Psi^{J_{A}}_{A}} || a^+_{\ell j}(\mathcal{B}) ||
  \Psi^{J_{A-1}}_{A-1} \rangle^2, \label{eq3}
\end{eqnarray}
where $a^+_{\ell j} (\mathcal{B})$ is a
creation operator associated with  a s.p. Berggren state $|\mathcal{B}\rangle$.
Since Eqs.~(\ref{ovf_GSM},\ref{eq3}) involve
summation  over all discrete Gamow states and integration over all 
scattering states along  the contour $L_+^{\ell_j}$, the final result is
independent of the s.p.~basis assumed. This is in contrast to standard
SF experimental extraction and SM calculations where model-dependence
enters through the specific choice of a s.p.~state $a^+_{n \ell j}$,
with Eq.~(\ref{eq3}) reducing to the sole matrix element
$\displaystyle \langle \Psi^{J_{A}}_{A} || a^+_{n \ell j} || 
\Psi^{J_{A-1}}_{A-1} \rangle^2$. 
This can lead to sizeable errors if the  states of $A-1$
and/or $A$  lie close to a channel threshold.

\begin{figure}[htb]
\centerline{\includegraphics[width=8.5cm,angle=0]{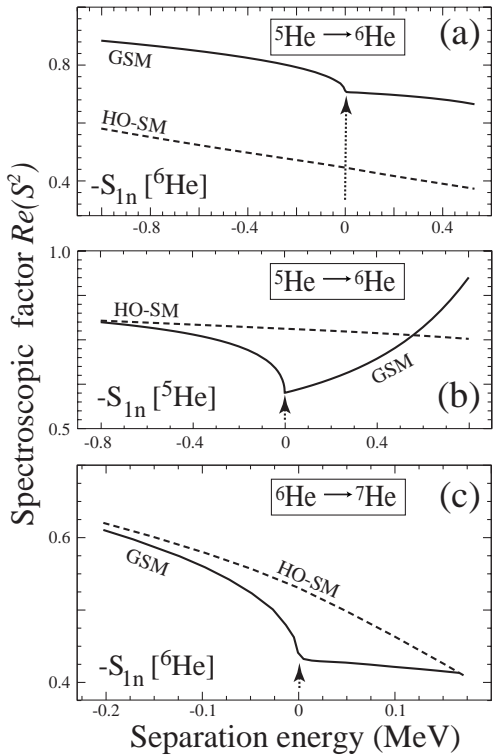}}
\caption{The real part of  the 
overlap integral as a function of one-neutron
separation energy  $S_{1n}$ as indicated.
Top and middle: 
$\displaystyle \langle ^6{\rm He( g.s.}) | 
[^5{\rm He( g.s.}) \otimes p_{3/2}]^{0^+} \rangle$; 
bottom:  $\displaystyle \langle ^7{\rm He( g.s.}) | 
[^6{\rm He( g.s.}) \otimes p_{3/2}]^{0^+} \rangle$.
Solid line:  GSM results; dashed line:  
the SM-like approximation (HO-SM) where the SGI matrix elements are calculated in 
the HO basis $\{ 0p_{3/2}, 0p_{1/2} \}$.  The 1n thresholds in $^6$He (top and
bottom) and $^5$He are marked by arrows.
The results displayed in the middle panel are plotted as a function of
the energy of the $0p_{3/2}$ resonant state, which, in our model,
is negative of $S_{1n}$[$^5$He].
For more details, see the discussion in the 
text.}
\label{fig_SF}
\end{figure}

\paragraph*{The Threshold Cusp--}
The GSM results for $^6$He g.s.~SF in the channel: $[{^5}{\rm He}({\rm
g.s.})\otimes p_{3/2}]^{0^+}$ are shown in Fig.~\ref{fig_SF}a as a
function of one-neutron separation energy  $S_{1n}$ in $^6$He. To this
end, we have fixed the depth of the  WS potential so that $0p_{3/2}$ and
$0p_{1/2}$ are both bound with respective energies -5 and -0.255 MeV,
and we have varied the SGI monopole coupling constant $V_0^{(J=0)}$ so
that the 1n separation energy of $^6$He goes through zero. At $S_{1n}$=0
(1n-emission threshold), the calculated SF exhibits behavior consistent
with the Wigner threshold  law: the quickly varying component of SF
behaves as $(-S_{1n})^{\ell-1/2}$ ($\ell$=1) {\it below} the 1n
threshold and follows the   $(S_{1n})^{\ell+1/2}$ rule   {\it above} the
 threshold. It is worth noting that the parameters of the GSM
Hamiltonian and the associated S-matrix poles do not show any
discontinuities around $S_{1n}$=0. This result constitutes an excellent
test of the GSM formalism: the Wigner limit is reached precisely at the
1n threshold obtained from many-body calculations. It is interesting to
see that the HO-SM results are depressed by about 25\% as compared to
GSM. Indeed, in the HO-SM  s.p. basis, the configurations $[0p_{3/2}
\otimes 0p_{3/2}]^{0^+}$ and $[0p_{1/2} \otimes 0p_{1/2}]^{0^+}$ are
strongly mixed by the SGI residual interaction; hence, the value of
$p_{3/2}$ SF in the g.s.~of $^6$He  is significantly reduced. Moreover,
the s.p.~basis in HO-SM calculations comprises HO states whose radial
form factors are independent of the depth of the WS potential. Hence, no
threshold effect can be seen in HO-SM SFs.

\paragraph*{Threshold Effects Due to Channel Coupling--}

As discussed above, when a new channel opens at energy $E_t$, there
appears a  flux redistribution in other open channels with lower
thresholds. This flux redistribution may be affected by the presence of
the Wigner cusp in the new channel; hence, the reaction cross-sections
in {\it all}   open channels may exhibit the threshold anomaly at $E_t$.
To illustrate the phenomenon of channel coupling, Fig.~\ref{fig_SF}b
shows again the SF for the  $[{^5}{\rm He}({\rm g.s.})\otimes
p_{3/2}]^{0^+}$ channel but this time as a function of the $0p_{3/2}$
s.p.~state energy (or negative $S_{1n}$ of $^5$He). This calculation was
carried out  by varying the depth of the WS potential so that the
$p_{3/2}$ pole of the $S$-matrix (which is also the g.s.~of $^5$He in
our model space), would change its character from a bound state to an
unbound decaying Gamow state. At $S_{1n}$[$^5$He]=0, the 1n and 2n
thresholds in $^6$He become degenerate, i.e., 
the two channels of $^5$He+n, namely $^6$He and $^4$He+2n couple.
As seen in Fig.~\ref{fig_SF}, the  SF in the $[{^5}{\rm He}({\rm
g.s.})\otimes p_{3/2}]^{0^+}$ channel exhibits the Wigner cusp at
$S_{1n}$[$^5$He]=0, i.e., at a 2n threshold. At this point, the first
derivative $\partial S^2/\partial e_{0p_{3/2}}$ of the SF becomes
infinite, consistent with the Wigner law for $\ell$=1. The SF in HO-SM
varies very little in the whole studied range of $0p_{3/2}$ energies.
(Let us mention that  the competition between bound and unbound 
states in $^6$He, and the $^5$He+n and $^4$He+2n reaction channels has been 
studied within the coupled cluster method \cite{Myo}.)

In the case shown in  Fig.~\ref{fig_SF}b the GSM Hamiltonian changes at
$S_{1n}$[$^5$He]=0 as at this point the $0p_{3/2}$ pole becomes unbound.
It is, therefore, instructive to investigate the situation at which the
Hamiltonian behaves smoothly around the threshold. Figure~\ref{fig_SF}c
illustrates  the case of $^7$He g.s.  in the channel $[{^6}{\rm He}({\rm
g.s.})\otimes p_{3/2}]^{3/2^-}$. In order to avoid  secondary
open-channel mixing effects, we have adjusted the WS potential depth so
that the $^5$He g.s. is bound by 5\,MeV and cannot play any role in the
anomalous energy dependence of the studied SF. The $0p_{1/2}$ state is
weakly  bound with an energy of $-0.255$ MeV. In order to control the 
binding energy of $^6$He, the $V_0^{(J=0)}$ coupling constant is varied
so that $^7$He g.s. is always bound while $S_{1n}$  of $^6$He goes
through zero. The GSM space is the same as in the previous cases, except
that the $0p_{1/2}$ state is now  included in the GSM basis as it is
bound. A cusp in the calculated SF for $^7$He is clearly seen at
$S_{1n}$[$^6$He]=0. The threshold anomaly shown in Fig.~\ref{fig_SF}c
can only result from the cross-channel couplings. Again, the   SF in 
HO-SM  smoothly varies  in the whole   energy region considered. 

\paragraph*{Conclusions--} The anomalies in the spectroscopic factors
when the total energy of the system varies through the threshold of an
opening channel are discussed within the many-body OQS formalism of the
GSM. The main conclusions of this work can be summarized as follows: (i)
Many-body OQS calculations correctly predict the Wigner-cusp and
channel-coupling threshold effects. This constitutes a very strong
theoretical check  for the  GSM approach; (ii) The spectroscopic
factors  defined in the OQS framework through the norm of the overlap
integral, exhibit strong variations around particle thresholds. Such
variations cannot be described in a standard CQS SM framework that
applies a `one-isolated-state' ansatz and ignores the coupling to the
decay and scattering channels. In our model calculations, the
contribution to SF from a non-resonant continuum can be as large as 25\%;
(iii) Any theoretical model aiming at a meaningful description of SFs of
low-$\ell$ states in weakly bound nuclei must meet certain minimal
conditions. Namely, it should be able to account for (a)  many-body
configuration mixing and the resulting spreading of the spectroscopic
strength due to inter-nucleon correlations, (b) coupling to the particle
continuum that affects radial properties of wave functions in the
neighborhood of reaction thresholds, and (c) coupling between various 
reaction channels.

Considering  points (ii) and (iii) above, one should be careful
when extracting spectroscopic information from transfer experiments  on
drip-line nuclei. The results presented
in this work suggest that, similar to other fields, experimental studies
of various aspects of   threshold effects  could provide valuable
spectroscopic information about the s.p. structure of weakly bound
nuclei. Here, of particular interest are systems  with near-lying 1n and
2n thresholds such as $^{6,8}$He or non-Borromean two-neutron halos
\cite{Timofeyuk03}, in which cusps in SFs are expected to be
particularly strong in low-$\ell$ ($\ell=0,1$) neutron channels. 
Similar features were found in the analysis of the continuum-coupling
correction to the CQS eigenvalues near the reaction threshold
\cite{Op05}. Finally, let us note that a threshold anomaly is also
expected  in proton-rich nuclei. While the effect is weaker than in the
neutron-rich systems, one still expects
anomalous  differences in SFs  and other spectroscopic quantities in
mirror nuclei. Work along these lines  is in progress.

This work was supported by  the U.S. Department of Energy
under Contract Nos. DE-FG02-96ER40963 (University of Tennessee),
DE-AC05-00OR22725 with UT-Battelle, LLC (Oak Ridge National
Laboratory), and DE-FG05-87ER40361 (Joint Institute for Heavy Ion
Research).

\end{document}